\documentclass[aps,prb,twocolumn,showpacs]{revtex4}
\usepackage{graphicx}
\usepackage{epsf}
\usepackage{amsmath}
\usepackage{amssymb}

\newcommand{\etal}{{\it et al.}}

\begin{document}

\title{Spin Hamiltonian of Hyperkagome Na$_4$Ir$_3$O$_8$}

\author{T. Micklitz} 

\affiliation{Dahlem Center for Complex Quantum Systems and 
Institut f\"ur Theoretische Physik,
Freie Universit\"at Berlin, 14195 Berlin, Germany}

\author{M. R. Norman} 

\affiliation{Materials Science Division, Argonne National Laboratory,
Argonne, Illinois 60439}

\date{\today} 


\begin{abstract}

We derive the spin Hamiltonian for the quantum spin liquid Na$_4$Ir$_3$O$_8$,
and then estimate the direct and superexchange contributions  between near
neighbor iridium ions using a tight binding parametrization of the electronic
structure.  We find a magnitude of the exchange interaction comparable to experiment for
a reasonable value of the on-site Coulomb repulsion.  For one of the two tight
binding parametrizations we have studied, the direct exchange term, which is
isotropic, dominates the total exchange.  This provides support for those theories proposed to describe
this novel quantum spin liquid that assume an isotropic Heisenberg model.
\end{abstract}
\maketitle

\section{Introduction}

Na$_4$Ir$_3$O$_8$ has captured much attention since
its discovery.\cite{okamoto,lee}  This insulator exhibits a large Curie Weiss temperature (650 K),
yet does not magnetically order down to the lowest measured temperature.
The reason for this is thought to be due to the strongly frustrated nature of the iridium lattice,
which forms a hyperkagome network composed of corner sharing triangles
(the hyperkagome lattice being formed by replacing one of the four Ir sites in a pyrochlore
lattice by a Na ion).
Most models that have been
proposed to describe this material assume an isotropic Heisenberg model with an effective
spin of 1/2.  This is
somewhat of a surprise, given the distorted nature of the lattice and the strong spin-orbit
coupling of the iridium ions.  These issues have been discussed in depth in Ref.~\onlinecite{chen}.

In a previous paper,\cite{norman} we have calculated the exchange constants based on a 
particular tight binding parametrization of the electronic structure, and found that the resulting spin 
Hamiltonian should be highly
anisotropic.  In the present paper, we revisit this issue by considering a more general tight binding
parametrization.
We now include the residual crystal field splittings of the Ir 5d orbitals due to the octahedral 
distortions, and find that 
this corrects a major deficiency 
of the previous fit, which was an anomalously large value for the $t_{dd}^\delta$ hopping.
As a result, the more general parametrization  leads to the exchange interaction being dominated 
by the direct exchange between Ir ions, and as a consequence, we find an approximately isotropic 
Heisenberg model.  We also find the observed magnitude of the exchange for a reasonable value
of the Coulomb repulsion.

In Section II, we provide a microscopic derivation of the exchange Hamiltonian,
finding a few differences from previously published results. 
In Section III, we describe the tight binding
parametrizations of the electronic structure,
and then in Section IV, the resulting exchange constants as a function of
the Coulomb repulsion.  In Section V, we summarize our findings.

\section{Exchange Couplings}

In Na$_4$Ir$_3$O$_8$, the $t_{2g}$ 5d manifold contains one hole per iridium site (the $e_g$
levels are empty), and the oxygen 2p levels are filled.~\cite{chen,norman}  This single hole
sits in a half-filled doublet level due to spin-orbit coupling, thus motivating an effective
S=1/2 exchange model.  This picture, which is supported by our electronic structure calculations,
can be exploited to calculate the exchange couplings following Ref.~\onlinecite{chen}.

\subsection{Microscopic Hamiltonian}

In our own derivation, we exploit  
 the work of Refs.~\onlinecite{YHAW,aharony}, which calculated
the exchange couplings for cuprates in the presence of spin-orbit coupling, following the earlier
work of Koshibae \etal~\cite{koshibae} referred to in Ref.~\onlinecite{chen}.
We want to derive the exchange couplings generated to fourth order in the hopping of Ir holes in the $t_{2g}$ 
complex. Hopping can be direct between the Ir$^{4+}$ ions or via the p-orbitals of the O$^{2-}$ ions.
We denote 
by $\epsilon^d_{1,2,3}$ the Kramers-degenerate energy levels of the $t_{2g}$ complex with 
$\epsilon^d_3>\epsilon^d_1,\epsilon^d_2$.  These splittings are mostly due to the spin-orbit coupling,
which acts to form a lower quartet and an upper doublet, but there is a contribution as well
from residual crystal field splittings resulting from the low site symmetry of the
iridium ion (noting that the space group is cubic).~\cite{chen,norman}  The vacuum state is denoted
as $|\Omega \rangle$ where all three levels are
fully occupied. The ground state is then generated by linear combinations of the form
\begin{align}
|\phi_0 \rangle 
= {\rm span}\{c^\dagger_{i3\uparrow} |\Omega \rangle, c^\dagger_{i3 \downarrow} |\Omega \rangle\}
\end{align} where $\uparrow,\downarrow$ characterizes the Kramers degenerate states, $i$
is the site index, and  $c^\dagger$ is the creation operator for a hole in the $t_{2g}$ complex.
According to the Goodenough-Kanamori rules, the strongest contribution to the exchange coupling 
results from two half-occupied orbitals, so we may focus only on the hopping between 
the $\epsilon^d_3$ orbitals. The Hamiltonian acting on the ground state is then
\begin{align}
H= H_0 + H_{\rm hop}
\end{align} 
with on-site Hamiltonian
\begin{align}
H_0 =& 
\sum_{im\sigma}\epsilon^d_m c^\dagger_{im\sigma} c_{im\sigma}
+ {U_d\over 2} \sum_{imn\sigma\sigma'} 
c^\dagger_{im\sigma}c^\dagger_{in\sigma'} c_{in\sigma'}c_{im\sigma} \nonumber \\
& +
\sum_{ka\sigma} \epsilon^p_{k a} p^\dagger_{k a\sigma}p_{ka\sigma} 
+ {U_p\over 2} \sum_{kab\sigma\sigma'} 
p^\dagger_{ka\sigma}p^\dagger_{kb\sigma'} p_{kb\sigma'}p_{ka\sigma}
\end{align} 
where $c^\dagger$ ($c$) and $p^\dagger$ ($p$) are creation
(annihilation) operators for holes. 
Indices $i$ and $k$ describe sites of the Ir ion and its nearest-neighbor 
O ions, $m,n$ and $a,b$ denote the Ir $t_{2g}$ and O $p$ levels, and $\sigma,\sigma'$
are spin indices. The relevant hopping between nearest neighbor Ir  and O sites
is 
\begin{align}
H_{\rm hop} = 
&
\sum_{ij\sigma\sigma'} 
(t^{ij}_{33})_{\sigma\sigma'} c^\dagger_{i3\sigma}c_{j3\sigma'} 
\nonumber \\
& +
\sum_{ika\sigma\sigma'}
\left(
 (t^{ki}_{a3})_{\sigma\sigma'} 
p^\dagger_{ka\sigma}c_{i3\sigma'} 
+ \text{h.c.}
\right)
\end{align} where the spin dependent matrix elements for hopping between Ir, and 
Ir and O ions, are
\begin{align}
(t^{ij}_{33})_{\sigma\sigma'} 
&= \tilde{t}^{ij}_{33} \delta_{\sigma\sigma'} 
+ \bold{C}^{ij}_{33} \cdot \bold{\sigma}_{\sigma\sigma'} \\
\left(t^{ki}_{a3}\right)_{\sigma\sigma'} 
&= \tilde{t}^{ki}_{a3} \delta_{\sigma\sigma'} 
+ \bold{C}^{ki}_{a3} \cdot \bold{\sigma}_{\sigma\sigma'}
\end{align}
Note that we use a formulation in terms of holes. But to compare with the results 
of Ref.~\onlinecite{chen}, we will insert
for the energies of virtual states that of the electrons and not the holes.

\subsection{Exchange couplings from perturbation theory}

The Kramers degenerate ground state is split due to virtual hopping processes. An effective spin 
Hamiltonian $H_S$ can be derived from perturbation theory in $H_{\rm hop}$. 
Accounting only for interactions between neighboring spins it
is of the general form
\begin{align}
H_S = \sum_{\langle ij \rangle} {\cal H}_{i,j}
\end{align} Here $\langle ij \rangle$ indicates a sum over nearest neighbor sites and 
\begin{align}
{\cal H}_{i,j} = \sum_{pq} J_{pq}(i,j)S_p(i)S_q(j)
\end{align} with spin-$1/2$ operators $\bold{S}$ ($p,q=x,y,z$). In what follows we want to 
derive expressions for $J_{pq}(i,j)$ up to
fourth order in the hopping. For this it is convenient to introduce
\begin{align} 
T^{dd}_{ji}
&= \sum_{\sigma\sigma'} 
(t^{ji}_{33})_{\sigma\sigma'} c^\dagger_{j3\sigma}c_{i3\sigma'}
\\
T^{pd}_{ki}
&=
\sum_{a\sigma\sigma'} 
(t^{ki}_{a3})_{\sigma\sigma'} 
p^\dagger_{ka\sigma}c_{i3\sigma'} 
\\
(T^{pd})^\dagger_{ik}
&=
\sum_{a\sigma\sigma'} 
(t^{ik}_{3a})_{\sigma\sigma'} 
c^\dagger_{i3\sigma}p_{ka\sigma'}
\end{align} so that
\begin{align}
H_{\rm hop} 
=&
 \sum_{ji}  
T^{dd}_{ji} 
 + \sum_{ki} \left( T^{pd}_{ki} + (T^{pd})^\dagger_{ik} \right) 
\end{align}
The energy of the ground state $|\phi_0\rangle$ is set to zero.

\subsubsection{Direct exchange}

The lowest order contribution results from direct hopping between Ir ions. 
There are two contributions resulting from the back and forth hopping of 
 holes on Ir sites $i$ and $j$, respectively. Both processes give identical contributions, so we can restrict
 ourselves to the hopping of a hole at site $i$ and multiply its contribution by a factor of two,  
 \begin{align}
{\cal H}^{(2)}_{i,j} 
=& - 2 \langle \phi_0 |  T^{dd}_{ij}{1\over H_0}  T^{dd}_{ji}| \phi_0 \rangle
\end{align} 
As mentioned above, we want to compare with the findings of Ref.~\onlinecite{chen}
and therefore we give the energy of the intermediate state with two holes on Ir site  $j$ in 
terms of the electron's energy. The latter is $-4U_d+5U_d=U_d$ and 
results from the difference in reduced (increased) Coulomb interaction 
on the site where a hole is inserted (removed). 
Inserting $T^{dd}_{ji}$ from Eq.~9 we get 
\begin{align}
{\cal H}^{(2)}_{i,j} 
= - {2\over U_d}  \sum_{\sigma_1\sigma_2\sigma_3\sigma_4} &
(t^{ij}_{33})_{\sigma_4\sigma_3} 
(t^{ji}_{33})_{\sigma_2\sigma_1} \nonumber \\
 &   
\langle \phi_0 | 
c^\dagger_{i3\sigma_4}c_{j3\sigma_3}
c^\dagger_{j3\sigma_2}c_{i3\sigma_1}
| \phi_0 \rangle   
\end{align} 
We now employ the identity 
(
for notational convenience we suppress the unit matrix in 
$\tfrac{1}{2}\openone$)
\begin{align}
c^\dagger_{i3\sigma_1}c_{i3\sigma_2} 
= \left( \tfrac{1}{2} + \bold{S}_i \cdot \bold{\sigma} \right)_{\sigma_2\sigma_1}
\end{align} to rewrite this as
\begin{align}
{\cal H}^{(2)}_{i,j} 
=& {2\over U_d} {\rm tr} \left(
t^{ij}_{33} \left( \tfrac{1}{2} + \bold{S}_j \cdot \bold{\sigma}  \right) t^{ji}_{33}
\left( \tfrac{1}{2} + \bold{S}_i \cdot \bold{\sigma}  \right)
\right)  
\end{align} where the trace is over spin-indices and we neglected a quadratic 
term  $cc^\dagger$ which does not contribute to the effective spin Hamiltonian in Eq.~8. 
Terms involving the factor $1/2$ 
lead to contributions which also do not
 contribute to Eq.~8. 
 Neglecting these 
 we find
\begin{align}
{\cal H}^{(2)}_{i,j} 
=& {2\over U_d} {\rm tr} \left(
t^{ij}_{33} (\bold{S}_j \cdot \bold{\sigma}) t^{ji}_{33} (\bold{S}_i \cdot \bold{\sigma})
\right)   
\end{align}
This expression is a factor of two larger than that quoted in Ref.~\onlinecite{chen}.

\subsubsection{Superexchange via oxygen}

We now turn to processes involving intermediate oxygen and states arising
from fourth order hopping. Let us start with two observations.  First, only those processes where
the hole hops from one Ir via the O to a different Ir give spin dependent contributions, i.e.
processes where the hole hops to O and then returns to the same Ir it started from do not 
contribute to Eq.~8 and will be neglected in the following. Second, following Ref.~\onlinecite{aharony},
we notice that there are two qualitatively 
different processes in which fourth order hopping contributes to Eq.~8. 
In the `consecutive channel' (cc) a hole hops from an Ir ion  to the O, then to the second Ir, 
and finally back to the first Ir via the same  or a different  O. In this channel there are two holes on an Ir ion
in the intermediate state.
The consecutive channel cc gives an identical contribution as the
direct term, however with an effective hopping amplitude 
$t^{ij}_{33}\rightarrow \tau^{ij}_{33}$. 
In the second `simultaneous channel' (sc)  a hole hops from one of the Ir ions to the O, and then 
a second hole hops  from the second Ir to the same or a different O. Afterwards the holes return
to the Ir, i.e. back to the ground state in which one hole occupies each Ir. In this second
channel, two holes simultaneously occupy O as intermediate states.

\paragraph{Consecutive channel cc:}

Let us first turn to contributions from the consecutive channel (cc) and look at a hole at Ir site $i$. 
Hopping of the hole from site $i$ to $j$ via an  
O ion at site $k$ results in a state $|\phi^{cc}_{2j} \rangle$ with two holes occupying 
$t_{2g}$ levels of Ir at site $j$ and no hole on Ir site $i$, 
\begin{align}
H_0 |\phi^{cc}_{2j} \rangle
=& (T^{pd})^\dagger_{jk}{1\over H_0}  T^{pd}_{ki} | \phi_0 \rangle   
\end{align} The (electron) energy of the intermediate state after the first hop is 
a sum of four contributions.  Removing the hole from the $t_{2g}$
level and inserting the hole in the  $a$-level of the O ion at site $k$
gives $\epsilon^d_3-\epsilon^p_{k a}$ for the on-site energies 
and $5(U_d - U_p)$ from the increased and reduced Coulomb interaction 
between the electrons on Ir and O ions, respectively. That is, the energy
of the intermediate state after the first hop is 
\begin{align}
\epsilon^{pd}_{ka} = \epsilon^d_3-\epsilon^p_{k a} + 5(U_d - U_p)
\end{align} Inserting Eq.~10 then gives
\begin{align}
H_0 |\phi^{cc}_{2j} \rangle
=& 
\sum_{ak\sigma_1\sigma_2\sigma_3} 
{1\over \epsilon^{pd}_{ka} }
(t^{jk}_{3a})_{\sigma_3\sigma_2}
\left(t^{ki}_{a3}\right)_{\sigma_2\sigma_1}  \nonumber \\
& \quad \times c^\dagger_{j3\sigma_3}p_{ka\sigma_2}
p^\dagger_{ka\sigma_2}c_{i3\sigma_1} | \phi_0 \rangle 
\nonumber \\
=& 
\sum_{ak\sigma_1\sigma_2\sigma_3}
{1\over \epsilon^{pd}_{ka} }
(t^{jk}_{3a})_{\sigma_3\sigma_2}
\left(t^{ki}_{a3}\right)_{\sigma_2\sigma_1}
c^\dagger_{j3\sigma_3} c_{i3\sigma_1} | \phi_0 \rangle 
\end{align} 
Introducing the effective hopping between Ir sites
\begin{align}
\tau^{ji}_{33}
=& 
\sum_{ak} 
{{t}^{jk}_{3a}
t^{ki}_{a3}
\over \epsilon^{pd}_{ka} }
\end{align} which is a matrix in spin space (summation over 
intermediate spins is implicit),
 this is rewritten as
 \begin{align}
H_0 |\phi^{cc}_{2j} \rangle
=& 
\sum_{ \sigma_1\sigma_2} 
(\tau^{ji}_{33})_{\sigma_2\sigma_1}
c^\dagger_{j3\sigma_2} c_{i3\sigma_1} | \phi_0 \rangle   
\end{align} Eq.~22 corresponds to the intermediate state
$T_{ji}^{dd} |\phi_0 \rangle$ in the direct exchange hole pathway Eq.~13  
with the hopping matrix elements $\tau^{ji}_{33}$. Therefore division by the 
(electron) energy $U_d$ of the intermediate state Eq.~22 and application of the second hop 
which returns the hole to Ir site $i$, we find
\begin{align}
{\cal H}^{(4 cc)}_{i,j} 
=& {2\over U_d} {\rm tr} \left(
\tau^{ij}_{33} (\bold{S}_j \cdot \bold{\sigma})
\tau^{ji}_{33} (\bold{S}_i \cdot \bold{\sigma})
\right)   
\end{align} where the factor of two accounts again for the fact that 
an identical contribution results from a process in which the hole 
starts at site $j$.

\paragraph{Simultaneous channel sc:} 

In the `simultaneous channel' sc both Ir-holes from sites $i$ and $j$ hop onto O ions 
in the intermediate state $|\phi^{sc}_{2}\rangle$. This can occur in a process
in which the first hole at site $i$ and then the hole at site $j$ hops, or in the reversed order.
The first hop in this process results in the state
\begin{align}
H_0 |\phi^{sc}_1\rangle 
&=
\sum_k \left( T^{pd}_{kj} +  T^{pd}_{ki} \right) |\phi_0\rangle 
\nonumber \\
& = 
\sum_{ak\sigma_1\sigma_2} 
\Big( 
(t^{ki}_{a3})_{\sigma_2\sigma_1} 
p^\dagger_{ka\sigma_2}c_{i3\sigma_1}
\nonumber \\ 
& \qquad \qquad \qquad +
(t^{kj}_{a3})_{\sigma_2\sigma_1} 
p^\dagger_{ka\sigma_2}c_{j3\sigma_1}
\Big)
|\phi_0\rangle
\end{align} The energy of the intermediate state $|\phi^{sc}_1\rangle$ is
again $\epsilon^{pd}_{ka}$ in Eq.~19. Adding then the second hole on the same or a different 
O ion, we obtain the intermediate state $H_0 |\phi^{sc}_{2} \rangle$, i.e.
\begin{align}
H_0|\phi^{sc}_{2} \rangle
&=
\sum_{lk} \left( T^{pd}_{li}  {1\over H_0}  T^{pd}_{kj} 
+ T^{pd}_{lj}  {1\over H_0}  T^{pd}_{ki} \right) |\phi_0\rangle 
\end{align} An explicit expression for this intermediate state is given in Appendix~A, and
we here only remark that its (electron) energy 
depends on whether both holes are on the same O ion or not.
It has the non-interacting contribution 
$2\epsilon^d_3 - \epsilon^p_{ka} - \epsilon^p_{lb}$,
an interaction contribution $5U_d+5U_d$ from removing holes at
Ir sites $i$ and $j$, and the interaction contribution $-9U_p$
if both holes are on the same O ion and $-10 U_p$ in case they
are not. This can be summarized as
\begin{align}
\epsilon^{pd}_{lb}+\epsilon^{pd}_{ka} + U_p\delta_{kl}
\end{align} The return to the ground state can occur again in two ways,
i.e., a hole returns first to Ir at site $i$ and then the second returns to 
site $j$, or in the reversed order. The overlap of the resulting state 
with the ground state defines the effective Hamiltonian
\begin{widetext}
\begin{align}
{\cal H}^{(4 sc)}_{i,j} = \sum_{nm} 
\langle \phi_0|
\left( 
(T^{pd})^\dagger_{in}  {1\over H_0}  (T^{pd})^\dagger_{jm}
+
(T^{pd})^\dagger_{jn}  {1\over H_0}  (T^{pd})^\dagger_{im}\right) 
 |\phi^{sc}_2\rangle
\end{align} 
\end{widetext}
The calculation of Eq.~27 is identical to the one for cuprates in Refs. 5,6. 
For completeness we give details in Appendix~B, and here merely state that 
neglecting spin independent contributions, the effective Hamiltonian of interest 
is
\begin{align}
{\cal H}^{(4 sc)}_{i,j} 
=&
\sum_{balk}
{1
\over 
\epsilon^{pd}_{lb}+\epsilon^{pd}_{ka} + U_p\delta_{kl}
}
\left( {1\over \epsilon^{pd}_{lb}} + {1\over \epsilon^{pd}_{ka}} \right)^2
\nonumber \\
&\quad \times
{\rm tr}\left(
t^{il}_{3b}
t^{lj}_{b3}
\left( \bold{S}_j  \cdot \bold{\sigma} \right)
t^{jk}_{3a}
t^{ki}_{a3}
\left(  \bold{S}_i  \cdot \bold{\sigma} \right)
\right)
\end{align} Finally, summing the superexchange contributions 
from both channels
the superexchange contribution is
\begin{align}
{\cal H}^{(4)}_{i,j} 
=& \sum_{balk}
g^{lk}_{ba}
{\rm tr}\left(
t^{il}_{3b}
t^{lj}_{b3}
\left( \bold{S}_j  \cdot \bold{\sigma} \right)
t^{jk}_{3a}
t^{ki}_{a3}
\left(  \bold{S}_i \cdot \bold{\sigma} \right)
\right)
\end{align} where we
introduced
\begin{align}
g^{lk}_{ba}
=& 
{2\over \epsilon^{pd}_{lb} \epsilon^{pd}_{ka}  U_d} 
+
{1
\over 
\epsilon^{pd}_{lb}+\epsilon^{pd}_{ka} + U_p\delta_{kl}
}
\left( {1\over \epsilon^{pd}_{lb}} + {1\over \epsilon^{pd}_{ka}} \right)^2
\end{align} 
Note that the expression for $g$ differs from that of Ref.~\onlinecite{chen}.
This difference was previously noted in Refs.~\onlinecite{YHAW, aharony, koshibae2}
in connection with the similar expression in Ref.~\onlinecite{koshibae}.

\subsection{Exchange Constants}

We next want to bring Eq.~17 
for the direct exchange into the form
\begin{align}
{\cal H}_{i,j}= J \bold{S}_i \cdot \bold{S}_j 
+\bold{D}^{ij} \cdot (\bold{S}_i \times \bold{S}_j)
+ \bold{S}_i \cdot \overset{\leftrightarrow}{\Gamma}^{ij}\cdot \bold{S}_j
\end{align} To this end we insert Eqs.~5 and 6 into Eq.~17
\begin{align}
{\cal H}^{(2)}_{i,j} 
=& {2\over U_d} {\rm tr} \left(
(\tilde{t}^{ij}_{33}+\bold{C}^{ij}_{33} \cdot \bold{\sigma} )
(\bold{S}_j \cdot \bold{\sigma}) 
(\tilde{t}^{ji}_{33}+\bold{C}^{ji}_{33} \cdot \bold{\sigma} ) 
(\bold{S}_i \cdot \bold{\sigma})
\right)  
\end{align} We then employ the 
identity
(checked in Appendix~B) 
\begin{align}
{1\over 2}&{\rm tr}  \left(  (A_1 +\bold{B}_1\cdot \bold{\sigma})(\bold{S}_i \cdot \bold{\sigma})
(A_2 +\bold{B}_2 \cdot \bold{\sigma})(\bold{S}_j \cdot \bold{\sigma}) \right) \nonumber \\
& =  
 A_1A_2   
\bold{S}_i \cdot  \bold{S}_j 
+
 i \left( 
 A_2\bold{B}_1  
-
A_1\bold{B}_2  \right) \cdot \left( \bold{S}_i\times \bold{S}_j \right) \nonumber \\
& \qquad +
 \bold{S}_i \cdot \left( 
 \overset{\leftarrow}{\bold{B}}_1  \overset{\rightarrow}{\bold{B}}_2  
 +
 \overset{\leftarrow}{\bold{B}}_2  \overset{\rightarrow}{\bold{B}}_1  
-  \bold{B}_1 \cdot \bold{B}_2 \overset{\leftrightarrow}{\openone} 
\right) 
\cdot \bold{S}_j
\end{align} 
to find that ${\cal H}^{(2)}_{i,j}$ is of the form Eq.~31 with
\begin{align}
J^{(2)} & = {4\over U_d} \tilde{t}^{ij}_{33} \tilde{t}^{ji}_{33}
\\
\bold{D}^{(2)}_{ij} &= -{4i \over U_d}  
\left( 
 \tilde{t}^{ji}_{33} \bold{C}^{ij}_{33} 
 -
\tilde{t}^{ij}_{33} \bold{C}^{ji}_{33}
\right) 
\\
\overset{\leftrightarrow}{\Gamma}^{(2)}_{ij}
&=
{4\over U_d}
\left(
\overset{ \leftarrow}{\bold{C}}^{ji}_{33}  
\overset{\rightarrow}{\bold{C}}^{ij}_{33}  
+
\overset{ \leftarrow}{\bold{C}}^{ij}_{33}  
\overset{\rightarrow}{\bold{C}}^{ji}_{33}  
-  
\bold{C}^{ij}_{33} \cdot \bold{C}^{ji}_{33}
\overset{\leftrightarrow}{\openone}
\right)
\end{align}
To do the same calculation 
for the superexchange Eq.~29, let us first rewrite products 
of hopping matrices
\begin{align}
t^{il}_{3b}
t^{lj}_{b3}
& =(\tilde{t}^{il}_{3b} + \bold{C}^{il}_{3b} \cdot \bold{\sigma})
(\tilde{t}^{lj}_{b3}+ \bold{C}^{lj}_{b3} \cdot \bold{\sigma}) \nonumber \\
&=
\tilde{t}^{il}_{3b} \tilde{t}^{lj}_{b3}
+
\bold{C}^{il}_{3b} \cdot \bold{C}^{lj}_{b3} \nonumber \\
& \qquad +
\left(
i (\bold{C}^{il}_{3b} \times \bold{C}^{lj}_{b3})
+
\tilde{t}^{il}_{3b}\bold{C}^{lj}_{b3}
+ \bold{C}^{il}_{3b} \tilde{t}^{lj}_{b3}
\right) \cdot \bold{\sigma}
\nonumber \\
&=
s_{ij}^{lb}
+
\bold{v}_{ij}^{lb} \cdot \bold{\sigma}
\end{align}
where in the last line we introduced
\begin{align}
s_{ij}^{lb}
&=
\tilde{t}^{il}_{3b} \tilde{t}^{lj}_{b3}
+
\bold{C}^{il}_{3b} \cdot \bold{C}^{lj}_{b3}\\
\bold{v}_{ij}^{lb}
&=
i (\bold{C}^{il}_{3b} \times \bold{C}^{lj}_{b3})
+
\tilde{t}^{il}_{3b}\bold{C}^{lj}_{b3}
+ \bold{C}^{il}_{3b} \tilde{t}^{lj}_{b3}
\end{align}
We can now apply again  Eq.~33 to find
\begin{align}
{\cal H}^{(4)}_{i,j} 
=& \sum_{balk}
g^{lk}_{ba}
{\rm tr}\Big(
(s_{ij}^{lb}+\bold{v}_{ij}^{lb} \cdot \bold{\sigma})
\left( \bold{S}_j  \cdot \bold{\sigma} \right)
\nonumber \\
&\qquad  \times
(s_{ji}^{ka} + \bold{v}_{ji}^{ka} \cdot \bold{\sigma} )  
\left(  \bold{S}_i  \cdot \bold{\sigma} \right)
\Big) \nonumber \\
= & 
J \bold{S}_i \cdot \bold{S}_j 
+\bold{D}^{ij} \cdot (\bold{S}_i \times \bold{S}_j)
+ \bold{S}_i \cdot \overset{\leftrightarrow}{\Gamma}^{ij} \cdot \bold{S}_j
\end{align}
with
\begin{align}
J^{(4)} 
& =  2\sum_{balk}
s_{ij}^{lb} g^{lk}_{ba}s_{ji}^{ka}
\\
\bold{D}^{(4)}_{ij} 
&= -2i \sum_{balk}\left(
\bold{v}_{ij}^{lb} g^{lk}_{ba}s_{ji}^{ka}
-
s_{ij}^{lb} g^{lk}_{ba}\bold{v}_{ji}^{ka}
\right)
\\
\overset{\leftrightarrow}{\Gamma}^{(4)}_{ij}
&= 2\sum_{balk}\left(
\overset{ \leftarrow}{\bold{v}}_{ij}^{lb}  
g^{lk}_{ba}
\overset{\rightarrow}{\bold{v}}_{ji}^{ka}  
+
\overset{ \leftarrow}{\bold{v}}_{ji}^{lb}  
g^{lk}_{ba}
\overset{\rightarrow}{\bold{v}}_{ij}^{ka} 
 -  \bold{v}_{ij}^{lb} g^{lk}_{ba} \cdot \bold{v}_{ji}^{ka}
\overset{\leftrightarrow}{\openone}
\right)
\end{align} and $g^{lk}_{ba}$ given in Eq.~30.

\section{Tight Binding Parametrization}

In Ref.~\onlinecite{norman}, we performed local density approximation calculations for
the electronic structure of Na$_4$Ir$_3$O$_8$, and then performed a tight binding fit
within a Slater-Koster formalism.  To understand the tight binding fit, we need to step
back and take a look at the electronic structure of this material.~\cite{chen,norman} 
Na$_4$Ir$_3$O$_8$ is composed of IrO$_6$ octahedra, and as with many transition
metal oxides, these octahedra are distorted, with just a C$_2$ symmetry axis preserved.
Moreover, there are two types of oxygens, with four of the six around an Ir ion being of
one type (O2), the other two of the other type (O1).  Note that the two O1 ions are not
related by an inversion like in cuprates, but are related by a $\pi$ rotation about
the C$_2$ axis (see Fig.~1).  A minimal tight binding model restricted to near neighbors
only would then consist of hoppings of Ir to O1, Ir to O2, O1 to O1, O2 to O2,
O1 to O2, and Ir to Ir.  Because of the distorted nature of the lattice, not all distances between
given atom types are the same.  To reduce the number of fit parameters, we then assumed
a typical inverse fourth power dependence of the hopping integrals with distance for a given
atom combination.  In practice, this affects only the O to O hoppings, and we note that
an inverse fourth power behavior was indeed found for IrO$_2$.\cite{mattheis}
The net result is that we need 18 tight binding parameters:
energies of $t_{2g}$ ($\epsilon_{t_{2g}}$) and $e_g$ ($\epsilon_{e_g}$)
5d orbitals on Ir, energies of 2p orbitals on O1 and O2 ($\epsilon_{O1}$, $\epsilon_{O2}$)
d-d hoppings ($t_{dd}^\sigma$, $t_{dd}^\pi$, $t_{dd}^\delta$),
d-p hoppings ($t_{dp}^\sigma$, $t_{dp}^\pi$ for Ir-O1 and Ir-O2),
p-p hoppings ($t_{pp}^\sigma$, $t_{pp}^\pi$ for O1-O1, O2-O2, O1-O2), and the coefficient of the spin
orbit splitting for the Ir 5d orbitals, $\lambda$ (i.e., $\lambda~{\bf l\cdot s}$).

These parameters are then used to evaluate the various elements of the secular matrix that
generates the eigenvalues.\cite{slater}
The diagonal elements are simply given by the various on-site energies, and the spin-orbit coupling
matrix elements in a crystal field basis compatible with that used in Ref.~\onlinecite{slater} can 
be found in Ref.~\onlinecite{jones}.
The off-diagonal matrix elements are of the form $t_{ai,bj}(l,m,n) e^{i {\bf k} \cdot ({\bf r}_j-{\bf r}_i)}$
where $t$ is the hopping integral between orbital $a$ on site $i$ and orbital $b$ on site $j$, and
$l,m,n$ are the three direction cosines between the two sites at ${\bf r}_i$ and ${\bf r}_j$.
In our case, the resulting secular matrix has dimension 312 (there are four formula units in the unit
cell, i.e., there are 120 Ir 5d orbitals and 192 O 2p orbitals in the unit cell when spin-orbit
is included).  In considering off-diagonal elements of the secular matrix, note that each iridium ion is surrounded by
six oxygens and four other iridium ions (Fig.~1), and each oxygen is surrounded by twelve other
oxygens.  Once the secular matrix is set up, then the various tight binding parameters are iteratively 
adjusted to achieve an optimal fit to the band structure eigenvalues.
The function being minimized is a sum
of the squares of the differences of the tight binding eigenvalues from the
first principles ones of the electronic structure calculation.
The minimization was performed using Powell's method.\cite{numerical}
We first fit the calculation without spin-orbit using 42 eigenvalues
(the bottom and tops of the O1, O2, and $e_g$
complexes, as well as all 36 bands of the $t_{2g}$ complex)
at each of the four symmetry points of the simple cubic Brillouin zone.
As an initial start to the minimization, we used previously derived tight binding parameters for 
IrO$_2$,\cite{mattheis}
scaled (by an assumed inverse fourth power dependence on distance) to the Na$_4$Ir$_3$O$_8$
lattice.  The starting values for the on-site energies were estimated from the orbital decomposed 
density of states of the band calculation.  After minimization,
the resulting tight binding parameters were then used as input to fitting the calculation with spin-orbit,
now involving 78 eigenvalues
(the bottom and tops of the O1, O2, and $e_g$
complexes, as well as all 72 bands of the $t_{2g}$ complex), again at the four symmetry points.
For the initial start, the spin-orbit coupling parameter, $\lambda$, was estimated from the splitting of
the $j=3/2$ and $j=5/2$ eigenvalues of the band calculation.  After minimization,
the resulting fit gave a good reproduction of the energy bands, along with the Fermi surface
(Figs.~8 and 9 of Ref.~\onlinecite{norman}).
This is non-trivial, given the low site symmetry of this lattice, the size of the secular matrix,
and the large 18 parameter function space.
\begin{figure}
\centerline{
\includegraphics[width=2.4in]{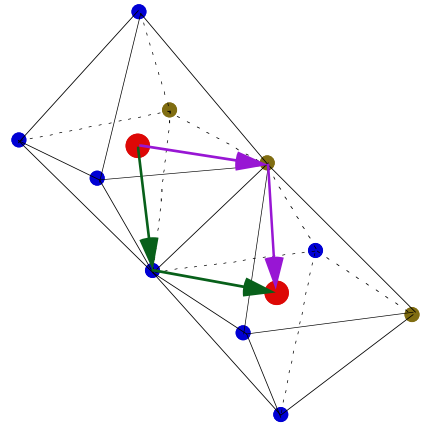}
}
\caption{(Color online) Superexchange pathway between two Ir ions as marked by the arrows.
The Ir ions are the large (red) spheres and the oxygen ions the small ones.  The lower left (green) 
pathway is via an O2 (blue) ion, the upper right (purple) pathway via an O1 (brown) ion.}
\label{fig1}
\end{figure}

This fit, though, does not take into account the residual crystal field splitting of the 5d orbitals
due to the distortions of the octahedra. These splittings, although somewhat obscured by
hybridization, appear to be present (Fig.~3 of Ref.~\onlinecite{norman}), and could potentially  be of
importance.
Including them increases the
number of tight binding parameters to 21 (with two more parameters needed for  the $t_{2g}$ manifold
and one for the $e_g$ one).  A complete description of these splittings are complicated because they involve a number of crystal field potential terms,
$V_{20}, V_{21}, V_{22}, V_{40}, V_{41}, V_{42}, V_{43}, V_{44}$, whose coefficients
are difficult to estimate from first principles. Rather, we used
a simplified approach where we ignore the small coupling between the $t_{2g}$ and $e_g$
orbitals.  For a C$_2$ axis along (1,1,0), the $t_{2g}$ diagonal elements of the secular matrix would
be (relative to $\epsilon_{t_{2g}}$) 
of the form -2$c_1$ for $xy$, and +$c_1$ for $xz$ and $yz$, with an off-diagonal matrix element 
$\pm c_3$ between $xz$ and $yz$, the sign depending on whether the C$_2$ axis  is along (1,1,0)
or (1,-1,0).  Diagonalization of this sub-matrix leads to two even symmetry states and one odd symmetry
state relative to the C$_2$ axis.  We note that this simplified
form ignores the smaller off-diagonal matrix element between $xy$ and $xz,yz$ which would couple
the two even symmetry states.
For the $e_g$ sub-matrix, the diagonal matrix elements (relative to $\epsilon_{e_g}$) 
are -2$c_2$ for $x^2-y^2$ and +2$c_2$ for $3z^2-r^2$.  Diagonalization of this sub-matrix leads
to one even symmetry and one odd symmetry state.
As the C$_2$ axis is rotated from one iridium site to the next, these $t_{2g}$ and $e_g$ 
sub-matrices in turn 
must be rotated (leading to off-diagonal terms between $x^2-y^2$ and $3z^2-r^2$).

The resulting 21 parameter tight binding fit had about a 25\% smaller RMS error than the 
18 parameter one
(the band dispersion from the fit is plotted in Fig.~2).  On the other
hand, the solution space is more complex than in the 18 parameter case, and as a consequence,
it is difficult to estimate how well the
minimization routine has succeeded in finding an optimal solution.  In particular, the $c_i$
parameters (Table I) lead to effective level splittings for the $t_{2g}$ and $e_g$ orbitals that do 
not seem to correspond 
well to those indicated by the band structure calculation (Fig.~3 of Ref.~\onlinecite{norman}).  Moreover,
the Fermi surface is somewhat
degraded relative to the one of the 18 parameter fit (Fig.~9 of Ref.~\onlinecite{norman}).
On the other hand, the 21 parameter
fit corrects a major deficiency of the 18 parameter one, in that $t_{dd}^\delta$, which was anomalously large in the 18 parameter fit
(0.1545 eV), is now far more reasonable (0.0521 eV). 
This, and other differences in the tight
binding parameters (comparing Table I with Table II of Ref.~\onlinecite{norman}), turn out to have a qualitative impact on the exchange constants, as we will see in the next Section.

\begin{table}
\caption{Tight binding hopping parameters in eV from the 21 parameter fit.  The on-site energies
are $\epsilon_{O1}$ = -6.4241,
$\epsilon_{O2}$ = -3.9141, $\epsilon_{t_{2g}}$ = -1.7230, $\epsilon_{e_{g}}$ = 0.6619,
with the spin-orbit coupling $\lambda$ = 0.5797.  The residual crystal field splittings of the cubic
levels on the Ir sites are denoted as $c_i$ ($i=1,2,3$).
}
\begin{ruledtabular}
\begin{tabular}{lccc}
 & $\sigma$ & $\pi$ & $\delta$ \\
\colrule
Ir-O1 &  -1.6015  & 0.8671 & \\
Ir-O2 &  -2.4604  & 1.1507 & \\
Ir-Ir & -0.4799  & 0.0049  & 0.0521 \\
O1-O1 & 0.5694  & 0.0284 & \\
O2-O2 & 0.4823 & -0.3264 &\\
O1-O2 & 0.6261  & 0.2560 & \\
\colrule
 & 1 & 2 & 3 \\
\colrule
$c_i$ & 0.0324 & -0.3839  & 0.2965 \\
\end{tabular}
\end{ruledtabular}
\end{table}

\begin{figure}
\centerline{
\includegraphics[width=3.4in]{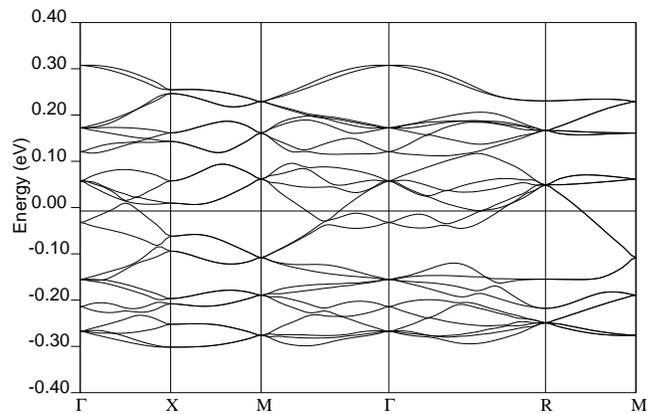}
}
\caption{Energy bands ($t_{2g}$ spin-orbit `doublet') near the Fermi energy 
($E_F$) from the 21 parameter tight binding fit to
the electronic structure of Na$_4$Ir$_3$O$_8$.  The horizontal line marks $E_F$.}
\label{fig2}
\end{figure}

\section{Exchange Constants}

The exchange constants are derived by taking the tight binding parameters discussed in the previous Section
and inserting them into the expressions derived in Section II (with $\epsilon^d_3=0$).
We remind that the functional form for the spin
Hamiltonian is listed in Eq.~31.  As stated previously in Ref.~\onlinecite{norman}, the direct exchange term
only contributes to the isotropic term $J$ in Eq.~31; the other terms vanish, even for the distorted lattice.
This is a consequence of the fact that the spin-orbit coupling leads to a `doublet' of states around
the Fermi energy (the 24 bands shown in Fig.~2) which to a good approximation are formed from linear 
combinations of $xy$, $xz$, and $yz$ orbitals with equal weights.
The resulting contribution to $J$
is of the form 4$t_d^2/U$ where $t_d =  \frac{1}{4}t_{dd}^{\sigma}+\frac{1}{3}t_{dd}^{\pi} 
+\frac{5}{12}t_{dd}^{\delta}$ 
(this differs by a factor of two from Ref.~\onlinecite{norman},
as already noted after Eq.~17).
  On the other hand, the superexchange terms contribute to $J$, ${\bold D}$
and $\overset{\leftrightarrow}{\Gamma}$.  We denote the components of the diagonal contributions
to the exchange by $J_i \equiv J + D_i + \Gamma_{ii}$ ($i = x, y, z$).  These values are plotted in Fig.~3
as a function of $U_d$ (Coulomb repulsion on the Ir sites) for both sets of tight binding parameters.  For
the purposes of these plots, $U_p$ (Coulomb repulsion on the O sites) was set to zero.

\begin{figure}
\centerline{
\includegraphics[width=3.4in]{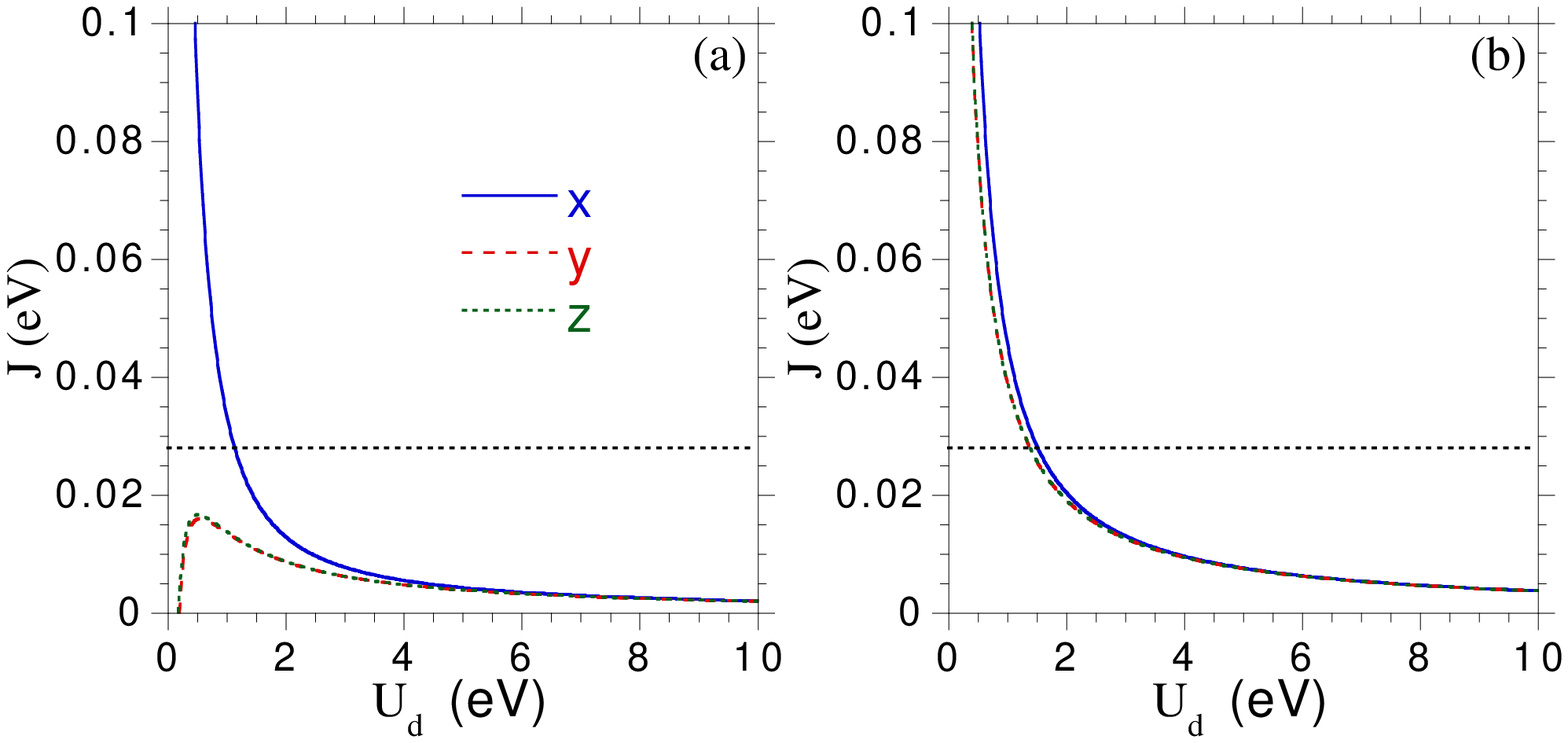}
}
\caption{(Color online) Diagonal elements of the exchange interaction as a function of the on-site repulsion ($U_d$)
on the iridium sites, assuming zero repulsion ($U_p$) on the oxygen sites;
(a) is from the 18 parameter tight binding fit and (b) from the 21 parameter fit.
The horizontal dashed line is the experimental value for $J$.~\cite{okamoto}}
\label{fig3}
\end{figure}

For the 18 parameter fit, $J_x$ becomes equal to the experimental value of 28 meV~\cite{okamoto}
for a value of $U_d$ of about 1.1 eV (Fig.~3a).  In that context, although the value of $U_d$ is not known
for Na$_4$Ir$_3$O$_8$, we note that for the related perovskite, Sr$_2$IrO$_4$, the optical gap is
0.5 eV, and it is known from LDA+U simulations that a $U_d$ of about 2 eV is needed to reproduce this
gap.\cite{jjyu}  Interestingly, $J_y$ and $J_z$ significantly differ from $J_x$, indicating that
the predicted spin Hamiltonian from this tight binding fit is strongly anisotropic, as we previously
remarked.\cite{norman}

\begin{table}[b]
\caption{Exchange constants in meV from the 21 parameter tight binding fit.
The spin Hamiltonian is given in Eq.~31.
Quoted are values where Ir site $m$ is along an (0,1,-1) direction relative to Ir site $n$.
$D$ are the Dzyaloshinski-Moriya, and
$\Gamma$ the anisotropic superexchange terms (in the second row for $\Gamma$,
the $jk$ refer to the parenthesis, $xy$, etc.).  In addition, the direct exchange (isotropic) is $J_d$ = 24.9, and
the isotropic superexchange term is $J_s$ = 2.2.  The last row, $J_{i}$, is the total exchange for the
diagonal components ($J_d + J_s + \Gamma_{ii}$).
The assumed value of $U_d$ is 1.5 eV (with $U_p$ assumed to be zero).
}
\begin{ruledtabular}
\begin{tabular}{lccc}
$i~(jk)$ & $x~(xy)$ & $y~(xz) $ & $z~(yz)$ \\
\colrule
D$_i$  & 3.4  & 0.4 & -0.3 \\
$\Gamma_{ii}$ & 1.3  & -1.3  & -1.3 \\
$\Gamma_{jk}$ & 0.3  & -0.2  & -0.0 \\
$J_i$ & 28.4  & 25.8  & 25.8 \\
\end{tabular}
\end{ruledtabular}
\end{table}

We can contrast this with the 21 parameter fit, shown in Fig.~3b.  $J_x$ reaches the experimental value of
28 meV for a $U_d$ of about 1.5 eV, which is close to the anticipated value of 2 eV.
More interestingly, $J_y$ and $J_z$ are close in value to $J_x$.  This indicates a far more isotropic spin Hamiltonian, which is in support of various theories for this material.  For completeness, we list all the
coefficients of the spin Hamiltonian in Table II for the 21 parameter fit with $U_d$ = 1.5 eV.  We note not only the
effective isotropy of $J_i$, but also the much reduced value of the 
Dzyaloshinski-Moriya interaction compared
to what was previously indicated in Ref.~\onlinecite{norman}.  This difference is mainly due to the much
smaller value of $U_d$ (0.5 eV) assumed in the previous work.
On general grounds, we note the dominance of $J_d$ in Table II compared to the other exchange
terms, in particular, the large ratio of $J_d$ to $J_s$.  This is in contrast to the well known
case of cuprates, where the superexchange term is dominant.  This difference can be attributed to
the 90 degree Ir-O-Ir bond present in this material compared to the 180 degree Cu-O-Cu bond found
in the cuprates.  Other qualitative differences between the 90 and 180 degree cases have been emphasized in the recent work of Jackeli and Khaliullin.\cite{jackeli}

Finally, we show in Fig.~4 the dependence of the exchange constants on $U_p$.  Its effect is to cause
increased anisotropy.  
However, this is more pronounced for the 18 parameter fit  than the 21 parameter one.
At larger values of $U_p$ than those shown in Fig.~4, there are divergences that
are associated with zeros
of the denominators entering $g$ (Eq.~30).  This corresponds to intermediate states that are low in energy,
meaning that perturbation theory is no longer valid.  Obviously, for very large values of $U_p$, the
superexchange contributions disappear, leaving only the isotropic direct exchange term.

\begin{figure}
\centerline{
\includegraphics[width=3.4in]{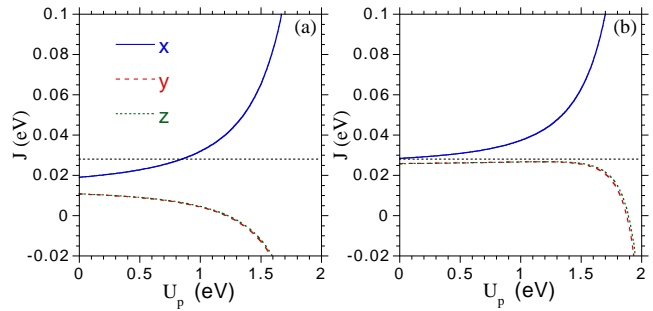}
}
\caption{(Color online) Diagonal elements of the exchange interaction as a function of the on-site repulsion ($U_p$)
on the oxygen sites, with a repulsion ($U_d$) of 1.5 eV on the iridium sites;
(a) is from the 18 parameter tight binding fit and (b) from the 21 parameter fit.
The horizontal dashed line is the experimental value for $J$.~\cite{okamoto}}
\label{fig4}
\end{figure}

From the above, it is obvious that the exchange constants are very sensitive to the tight binding
parametrization, and also the values of the various Coulomb repulsions.  
We also note that higher
order processes are ignored, for instance, sixth order superexchange processes involving additional hoppings
between the two O ions connecting two Ir sites which are known to play a role for the honeycomb lattice found
in Na$_2$IrO$_3$.~\cite{shitade}  Longer range hoppings on the iridium sublattice could also be
of importance as well.~\cite{podolsky}

\section{Summary}

We have found that an approximately isotropic Heisenberg model can be motivated from a tight
binding parametrization of the electronic structure of
Na$_4$Ir$_3$O$_8$, with the experimental value of $J$ reproduced
for a reasonable value of the Coulomb repulsion, $U_d$. 
To obtain this result, it was important to account for the residual crystal field splittings of the 
Ir 5d orbitals due to the octahedral distortions.
Our findings are obviously of some importance in regards
to models for this quantum spin liquid, since anisotropy acts to stabilize long range magnetic 
order.\cite{chen}
The large value of the exchange is of much interest, since a large $J$ appears to be associated
with the unusual properties of cuprates, including their d-wave superconductivity.~\cite{RMP}

\begin{acknowledgments}

We thank Gang Chen and Jaejun Yu for discussions.
Work at Argonne National Laboratory was supported by the U.S. DOE, Office of Science, under 
Contract  No.~DE-AC02-06CH11357.

\end{acknowledgments}

\appendix

\begin{widetext}

\section{Intermediate states in the superexchange pathway}

For self-containedness this Appendix  provides intermediate steps of the superexchange 
calculations similar to those found in Ref.~\onlinecite{YHAW}. 
The explicit expression for the intermediate state in the simultaneous channel 
of the superexchange pathway, Eq.~25, is a sum of the two contributions 
\begin{align}
H_0|\phi^{sc,a}_{2} \rangle
&=
\sum_{lk} T^{pd}_{lj}  {1\over H_0}  T^{pd}_{ki}  |\phi_0\rangle 
=
- \sum_{ba}\sum_{lk}\sum_{\sigma_4\sigma_3}\sum_{\sigma_2\sigma_1} 
{1\over \epsilon^{pd}_{ka}}
(t^{lj}_{b3})_{\sigma_4\sigma_3} 
(t^{ki}_{a3})_{\sigma_2\sigma_1} 
p^\dagger_{lb\sigma_4}p^\dagger_{ka\sigma_2}
c_{j3\sigma_3} c_{i3\sigma_1}
|\phi_0\rangle \\ 
H_0|\phi^{sc,b}_{2} \rangle
&=\sum_{lk} T^{pd}_{li}  {1\over H_0}  T^{pd}_{kj}  |\phi_0\rangle 
=
- \sum_{ba}\sum_{lk}\sum_{\sigma_4\sigma_3}\sum_{\sigma_2\sigma_1} 
{1\over \epsilon^{pd}_{ka}}
(t^{li}_{b3})_{\sigma_4\sigma_3} 
(t^{kj}_{a3})_{\sigma_2\sigma_1} 
p^\dagger_{ka\sigma_2}p^\dagger_{lb\sigma_4}
c_{j3\sigma_1}c_{i3\sigma_3} 
|\phi_0\rangle
\end{align} where the minus sign 
results from exchanging operators $p^\dagger$ and $c$. 
Upon relabeling indices $a\leftrightarrow b$, $k\leftrightarrow l$,
$\sigma_2\leftrightarrow \sigma_4$, and $\sigma_1\leftrightarrow \sigma_3$
we add up both contributions, resulting in the intermediate state
\begin{align}
H_0 |\phi^{sc}_{2} \rangle
&=
- \sum_{ba}\sum_{lk}\sum_{\sigma_4\sigma_3}\sum_{\sigma_2\sigma_1} 
\left( {1\over \epsilon^{pd}_{lb}} + {1\over \epsilon^{pd}_{ka}} \right)
(t^{lj}_{b3})_{\sigma_4\sigma_3} 
(t^{ki}_{a3})_{\sigma_2\sigma_1} 
p^\dagger_{lb\sigma_4}p^\dagger_{ka\sigma_2}
c_{j3\sigma_3} c_{i3\sigma_1}
|\phi_0\rangle
\end{align} and the state $|\phi^{sc}_{2} \rangle$ has the energy Eq.~26.
The return to the ground state can occur in the two processes 
described in Eq.~27. In the intermediate state 
$H_0 |\phi^{sc}_{3i}\rangle = \sum_{m} (T^{pd})^\dagger_{im}  |\phi^{sc}_2\rangle $ of
the first process one hole has returned to site $i$,
\begin{align}
H_0 |\phi^{sc}_{3i}\rangle 
&=
- \sum_{cba}\sum_{mlk}\sum_{\sigma_6\sigma_5\sigma_4\sigma_3}
\sum_{\sigma_2\sigma_1} 
{(t^{im}_{3c})_{\sigma_6\sigma_5}
(t^{lj}_{b3})_{\sigma_4\sigma_3} 
(t^{ki}_{a3})_{\sigma_2\sigma_1} 
\over \epsilon^{pd}_{lb}+\epsilon^{pd}_{ka} + U_p\delta_{kl}}
\left( {1\over \epsilon^{pd}_{lb}} + {1\over \epsilon^{pd}_{ka}} \right)
c^\dagger_{i3\sigma_6}p_{mc\sigma_5}
p^\dagger_{lb\sigma_4}p^\dagger_{ka\sigma_2}
c_{j3\sigma_3} c_{i3\sigma_1}
|\phi_0\rangle
\end{align} Using fermion anti-commutation relations 
and keeping only those contributions that are not
annihilated upon acting on the ground state,
$p_{mc\sigma_5}p^\dagger_{lb\sigma_4}p^\dagger_{ka\sigma_2}
=
\delta_{ml}\delta_{bc}\delta_{\sigma_5\sigma_4}p^\dagger_{ka\sigma_2}
-
\delta_{km}\delta_{ac}\delta_{\sigma_5\sigma_2}p^\dagger_{lb\sigma_4}$,
which inserted into Eq.~A4 shows that 
$H_0|\phi^{sc}_{3i}\rangle
=H_0\left( |\phi^{sc,a}_{3i}\rangle+|\phi^{sc,b}_{3i}\rangle \right)$ 
where
\begin{align}
H_0 |\phi^{sc,a}_{3i}\rangle 
&=
 \sum_{ba}\sum_{lk}\sum_{\sigma_6\sigma_4\sigma_3}\sum_{\sigma_2\sigma_1} 
{(t^{il}_{3b})_{\sigma_6\sigma_4}
(t^{lj}_{b3})_{\sigma_4\sigma_3} 
(t^{ki}_{a3})_{\sigma_2\sigma_1} 
\over \epsilon^{pd}_{lb}+\epsilon^{pd}_{ka} + U_p\delta_{kl}}
\left( {1\over \epsilon^{pd}_{lb}} + {1\over \epsilon^{pd}_{ka}} \right)
p^\dagger_{ka\sigma_2}
c^\dagger_{i3\sigma_6}
c_{j3\sigma_3} c_{i3\sigma_1}
|\phi_0\rangle \\ 
H_0 |\phi^{sc,b}_{3i}\rangle 
&=-
 \sum_{ba}\sum_{lk}\sum_{\sigma_6\sigma_4\sigma_3}\sum_{\sigma_2\sigma_1} 
{(t^{ik}_{3a})_{\sigma_6\sigma_2}
(t^{lj}_{b3})_{\sigma_4\sigma_3} 
(t^{ki}_{a3})_{\sigma_2\sigma_1} 
\over \epsilon^{pd}_{lb}+\epsilon^{pd}_{ka} + U_p\delta_{kl}}
\left( {1\over \epsilon^{pd}_{lb}} + {1\over \epsilon^{pd}_{ka}} \right)
p^\dagger_{lb\sigma_4}
c^\dagger_{i3\sigma_6}
c_{j3\sigma_3} c_{i3\sigma_1}
|\phi_0\rangle
\end{align} States $|\phi^{sc,a}_{3i}\rangle$ and $|\phi^{sc,b}_{3j}\rangle$ 
have the energies
$\epsilon^{pd}_{ka}$ and $\epsilon^{pd}_{lb}$, respectively. Acting with a second 
hopping Hamiltonian the second hole returns to its ground state
\begin{align}
\sum_{m} &(T^{pd})^\dagger_{jm}  |\phi^{sc,a}_{3i}\rangle 
=
 \sum_{ba}\sum_{lk}
 \sum_{\sigma_8\sigma_6\sigma_4\sigma_3}\sum_{\sigma_2\sigma_1} 
{(t^{jk}_{3a})_{\sigma_8\sigma_2}
(t^{il}_{3b})_{\sigma_6\sigma_4}
(t^{lj}_{b3})_{\sigma_4\sigma_3} 
(t^{ki}_{a3})_{\sigma_2\sigma_1} 
\over 
(\epsilon^{pd}_{lb}+\epsilon^{pd}_{ka} + U_p\delta_{kl})
\epsilon^{pd}_{ka}}
\left( {1\over \epsilon^{pd}_{lb}} + {1\over \epsilon^{pd}_{ka}} \right)
c^\dagger_{j3\sigma_8}
c^\dagger_{i3\sigma_6}
c_{j3\sigma_3} c_{i3\sigma_1}
|\phi_0\rangle \\
\sum_{m} &(T^{pd})^\dagger_{jm}  |\phi^{sc,b}_{3i}\rangle 
=
- \sum_{ba}\sum_{lk}\sum_{\sigma_8\sigma_6\sigma_4\sigma_3}\sum_{\sigma_2\sigma_1} 
{(t^{jl}_{3b})_{\sigma_8\sigma_4}
(t^{ik}_{3a})_{\sigma_6\sigma_2}
(t^{lj}_{b3})_{\sigma_4\sigma_3} 
(t^{ki}_{a3})_{\sigma_2\sigma_1} 
\over 
(\epsilon^{pd}_{lb}+\epsilon^{pd}_{ka} + U_p\delta_{kl})
\epsilon^{pd}_{lb}}
\left( {1\over \epsilon^{pd}_{lb}} + {1\over \epsilon^{pd}_{ka}} \right)
c^\dagger_{j3\sigma_8}
c^\dagger_{i3\sigma_6}
c_{j3\sigma_3} c_{i3\sigma_1}
|\phi_0\rangle
\end{align} Processes in which the holes return 
 to the ground state in reversed order
 have the intermediate state $|\phi^{sc}_{3j}\rangle$ in which
 Ir ion at site $j$ and $i$ are in the ground and excited states, respectively.
This intermediate state is again a sum of two 
 contributions which are taken to the ground state $|\phi_0\rangle$ upon
acting with the second hopping Hamiltonian. The final state corresponds to
Eqs.~A7 and A8 upon exchanging $i\leftrightarrow j$ in the first and second matrix elements 
 (i.e., those returning the holes to their ground states) and the $c^\dagger$'s.
Upon relabeling of spin indices these four contributions to the superexchange pathway from 
the simultaneous channel can be combined into a sum of two contributions, whose overlap
with the ground state is
\begin{align}
\langle \phi_0|
\Big(
&
\sum_{ba}\sum_{lk}
 \sum_{\sigma_8\sigma_6\sigma_4\sigma_3}\sum_{\sigma_2\sigma_1} 
{(t^{jk}_{3a})_{\sigma_8\sigma_2}
(t^{il}_{3b})_{\sigma_6\sigma_4}
(t^{lj}_{b3})_{\sigma_4\sigma_3} 
(t^{ki}_{a3})_{\sigma_2\sigma_1} 
\over 
\epsilon^{pd}_{lb}+\epsilon^{pd}_{ka} + U_p\delta_{kl}
}
\left( {1\over \epsilon^{pd}_{lb}} + {1\over \epsilon^{pd}_{ka}} \right)^2
c^\dagger_{j3\sigma_8}
c^\dagger_{i3\sigma_6}
c_{j3\sigma_3} 
c_{i3\sigma_1} 
\nonumber \\
&
+
 \sum_{ba}\sum_{lk}
 \sum_{\sigma_8\sigma_6\sigma_4\sigma_3}\sum_{\sigma_2\sigma_1} 
{(t^{ik}_{3a})_{\sigma_8\sigma_2}
(t^{jl}_{3b})_{\sigma_6\sigma_4}
(t^{lj}_{b3})_{\sigma_4\sigma_3} 
(t^{ki}_{a3})_{\sigma_2\sigma_1} 
\over 
\epsilon^{pd}_{lb}+\epsilon^{pd}_{ka} + U_p\delta_{kl}
}
\left( {1\over \epsilon^{pd}_{lb}} + {1\over \epsilon^{pd}_{ka}} \right)^2
c^\dagger_{i3\sigma_8}
c^\dagger_{j3\sigma_6}
c_{j3\sigma_3} 
c_{i3\sigma_1}
\Big)
|\phi_0\rangle
\end{align} Permuting some of the $c,c^\dagger$ and applying identity Eq.~15,
this can be summarized in the more compact form
\begin{align}
{\cal H}^{(4 sc)}_{i,j} 
=&
\sum_{ba}\sum_{lk}
{1
\over 
\epsilon^{pd}_{lb}+\epsilon^{pd}_{ka} + U_p\delta_{kl}
}
\left( {1\over \epsilon^{pd}_{lb}} + {1\over \epsilon^{pd}_{ka}} \right)^2
{\rm tr}\left(
t^{il}_{3b}
t^{lj}_{b3}
\left( \tfrac{1}{2} + \bold{S}_j  \cdot \bold{\sigma} \right)
t^{jk}_{3a}
t^{ki}_{a3}
\left( \tfrac{1}{2} + \bold{S}_i  \cdot \bold{\sigma} \right)
\right)
\nonumber \\
+
&
 \sum_{ba}\sum_{lk}
 {1
\over 
\epsilon^{pd}_{lb}+\epsilon^{pd}_{ka} + U_p\delta_{kl}
}
\left( {1\over \epsilon^{pd}_{lb}} + {1\over \epsilon^{pd}_{ka}} \right)^2
{\rm tr}\left(
t^{jl}_{3b}
t^{lj}_{b3} 
\left( \tfrac{1}{2} + \bold{S}_j  \cdot \bold{\sigma} \right)
\right)
{\rm tr}\left(
t^{ik}_{3a}
t^{ki}_{a3} 
\left( \tfrac{1}{2} + \bold{S}_i  \cdot \bold{\sigma} \right)
\right)
\end{align} where the trace is again in spin-space. 
The second term as well as contributions
from $1/2$ in the first term lead to spin independent 
contributions. Neglecting these, we arrive at the effective Hamiltonian 
Eq.~28.
\end{widetext}

\section{Traces over Pauli matrices}

Starting out from the expression
\begin{align}
{\rm tr}&  \left(  (A_1 +\bold{B}_1\cdot\bold{\sigma})(\bold{S}_i\cdot\bold{\sigma})
(A_2 +\bold{B}_2\cdot\bold{\sigma})(\bold{S}_j\cdot\bold{\sigma}) \right) \nonumber \\
&=  
S^k_iS^l_j A_1A_2 {\rm tr}\left( \sigma^k \sigma^l  \right) 
+
S^k_iS^m_j A_1B^l_2 
{\rm tr}\left( \sigma^k \sigma^l \sigma^m \right) 
\nonumber \\
& +
S^l_iS^m_j A_2B^k_1
{\rm tr}\left( \sigma^k \sigma^l \sigma^m \right) 
+
S^k_iS^m_j B^n_1B^l_2 
{\rm tr}\left( \sigma^k \sigma^l \sigma^m  \sigma^n  \right) 
\end{align} we use the identities
\begin{align}
{\rm tr}\left(  \sigma^k\sigma^l \right) &= 2\delta_{kl} \\
{\rm tr}\left(  \sigma^k\sigma^l \sigma^m \right) 
&= 2i \epsilon_{klm} \\
{\rm tr}\left(  \sigma^k\sigma^l \sigma^m\sigma^n \right) 
&=  2(\delta_{kl}\delta_{mn}  -\delta_{km}\delta_{ln} + \delta_{kn}\delta_{lm}) 
\end{align}
to find
\begin{align}
{1\over 2}{\rm tr}&  \left(  (A_1 +\bold{B}_1\cdot\bold{\sigma})(\bold{S}_i\cdot\bold{\sigma})
(A_2 +\bold{B}_2\cdot\bold{\sigma})(\bold{S}_j\cdot\bold{\sigma}) \right) \nonumber \\
&=  
 A_1A_2 
 \bold{S}_i  \cdot \bold{S}_j 
+
 i \left( 
 A_2\bold{B}_1  
-
A_1\bold{B}_2 \right)
\cdot \left( \bold{S}_i\times \bold{S}_j \right) \nonumber \\
& \quad +
 \bold{S}_i \cdot \left( 
 \overset{\leftarrow}{\bold{B}}_1  \overset{\rightarrow}{\bold{B}}_2  
 +
 \overset{\leftarrow}{\bold{B}}_2  \overset{\rightarrow}{\bold{B}}_1  
 -  \bold{B}_1 \cdot \bold{B}_2  \overset{\leftrightarrow}{\openone}
\right) 
\cdot \bold{S}_j
\end{align}

\end{document}